\documentclass[apl, amsmath, amssymb, amsfonts, reprint, aip]{revtex4-1}
\usepackage{graphicx}
\usepackage{float}
\usepackage{subfloat} 
\usepackage{caption}
\usepackage{subcaption}
\usepackage{dcolumn}
\usepackage{bm}
\usepackage{hyperref}
\usepackage{url}
\usepackage[utf8]{inputenc}
\usepackage[T1]{fontenc}
\usepackage{mathptmx}
\usepackage{etoolbox}

\makeatletter
\def\@email#1#2{%
 \endgroup
 \patchcmd{\titleblock@produce}
  {\frontmatter@RRAPformat}
  {\frontmatter@RRAPformat{\produce@RRAP{*#1\href{mailto:#2}{#2}}}\frontmatter@RRAPformat}
  {}{}
}%
\makeatother

\begin{document}

\preprint{AIP/123-QED}

\title{Experimental Demonstration of Turbulence-resistant Lidar via Quantum Entanglement}
\def\UMBC{Department of Physics, University of Maryland Baltimore County, Baltimore, Maryland 21250, USA}
\def\NGC{Northrop Grumman Corporation, Linthicum Heights, Maryland 21090, USA}
\author{Binod Joshi}
\email {binodj1@umbc.edu}
\affiliation{\UMBC}
\author{Michael M. Fitelson}
\affiliation{\NGC}
\author{Yanhua Shih}
\affiliation{\UMBC}

\begin{abstract}

We report a proof-of-principle experimental demonstration of a turbulence-resistant quantum Lidar system. As a key technology for sensing and ranging, Lidar has drawn considerable attention for a study from quantum perspective, in search of proven advantages complementary to the capabilities of conventional Lidar technologies. Environmental factors such as strong atmospheric turbulence can have detrimental effects on the performance of these systems. We demonstrate the possibility of turbulence-resistant operation of a quantum Lidar system via two-photon interference of entangled photon pairs. Additionally, the reported quantum Lidar also demonstrates the expected noise resistance. This study suggests a potential high precision timing-positioning technology operable under turbulence and noise.

\end{abstract}

\maketitle

Exploitation of quantum mechanical properties such as entanglement is an important avenue in the ongoing efforts to develop quantum communication schemes that offer significant advantage over the existing technologies based on classical systems. Such attempts have also been made in the field of Lidar (Light Detection and Ranging), which has been established as a well-known technique for remote sensing/ranging and other applications derived from it \citep{Slepyan,ReviewArticle}. Specifically, attempts have been made to address the issues of environmental noise and lossy medium affecting classical Lidar systems by proposing entangled photons for probing the target via the measurement of their correlations \citep{Lloyd, Shapiro, Lopaeva, England, Liu, Blakey, Zhao}. Such studies have also been extended to achieve quantum imaging \citep{Defienne, Gregory}. The impact of turbulence on quantum optical communication systems, in general, has also drawn attention for theoretical \citep{Jha, SemenovBell} and experimental \citep{Shekel, Capraro} inquiry; although a study specifically focusing in the search for a quantum advantage in Lidar systems under the influence of turbulence has not yet been reported. Our work is the first of this kind and demonstrates a turbulence-resistant operation of a Lidar system based on entanglement.

High accuracy in timing and positioning schemes has been demonstrated in systems based on the ``non-locality'' property of entangled states  \citep{Valencia}. In practice, though, any \textit{nonlocal} applications may face serious environmental effect \citep{Disorder, RarityNoise, RarityQKD}. Of these, turbulence is of particular importance for distant propagation of entangled quantum system through atmosphere, which is usually formalized as ``atmospheric quantum channel'' \citep{SemenovChannel}. As we know, in any classical optical measurements, the detrimental effect of turbulence is inevitable \citep{Tatarski, Owens, Buck, Clifford, Bobroff}. However, a quantum Lidar system that measures the two-photon interference induced correlation, in principle, is able to achieve turbulence-resistant results, i.e., any rapid phase variations along the optical path due to random changes in composition, density, length, index of refraction, or medium vibration, as well as presence of any noise, do not affect the temporal two-photon correlation and thus its precise timing and positioning. The robustness of two-photon interference of thermal state against turbulence has been demonstrated in a Young's double-slit interferometer \citep{Smith}. The experiment presented here exploits the working mechanism of a turbulence-resistant quantum Lidar system which relies on the measurement of two-photon interference induced second-order coherence function $G^{(2)}(t_1 - t_2)$ via photon counting of entangled signal-idler photon pairs produced from continuous wave (CW) pumped spontaneous parametric down conversion (SPDC) \cite{Klyshko, ShihBook}. In the reported experiment, both the signal photon and the idler photon are in the form of continuous waves.  Nevertheless the temporal correlation measurement of the biphoton at different distances has shown a narrow, sharp function of $t_1 - t_2$, invariant under significant level of atmospheric turbulence with uncertainty in the order of picoseconds, mainly determined by the resolution of the coincidence measurement device.
\begin{figure*}
\begin{minipage}[ht]{1.0\textwidth}
\centering
{\includegraphics[width = 0.7\linewidth]{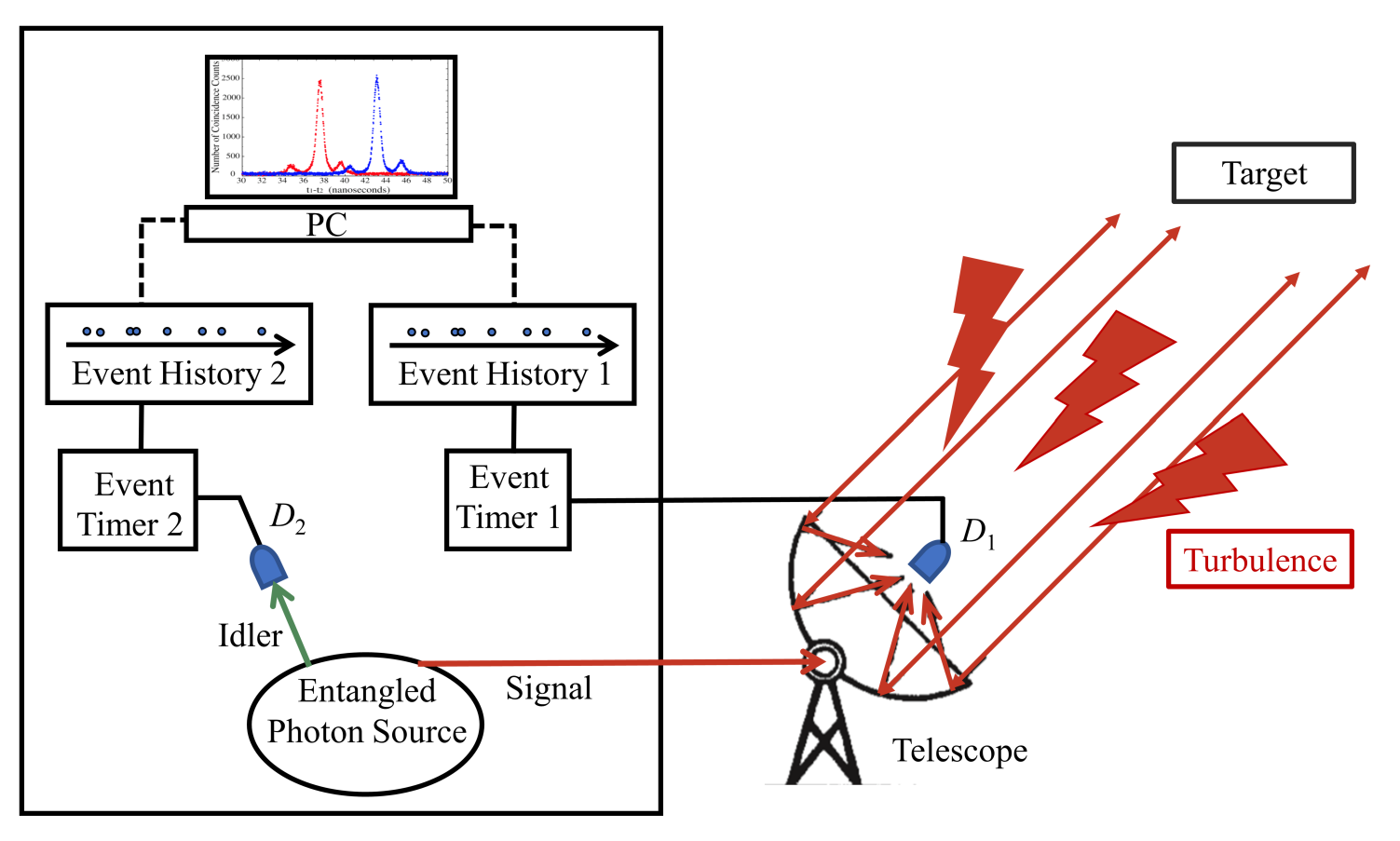}}
\end{minipage}
\caption{Schematic illustration of a Quantum Lidar system. }
\label{fig:schematics}
\end{figure*}

A classical measurement with time-correlated fast optical pulses may simulate a narrow correlation function in the absence of turbulence, thus making a basis for a high resolution Lidar system \citep{Murray, Wright}. The outcome of such measurement yields a convolution of two identical short pulses. However, when turbulence is added in the distant path of one or both pulses, the pulse(s) exposed to turbulence will be disturbed with temporal variations from time to time, hence resulting in a significantly broadened correlation function and a loss in the positioning/timing accuracy of the Lidar system. Additionally, when maximum usable power is limited, i.e., in weak light applications, conventional Lidar techniques based on laser pulses may not be suitable.  The Lidar scheme being reported in this work is designed to overcome these particular limitations of conventional Lidar systems, rather than a potential technology to substitute them. For weak light applications, when photon counting methods become necessary for correlation measurements, this Lidar scheme may be a more suitable choice \citep{RarityNoise}, especially when a turbulence-resistant operation is desirable.

A simple quantum Lidar setup consisting of CW pumped SPDC is illustrated in Fig.~\ref{fig:schematics}. The signal photon is sent to a distant target in a turbulent medium at position $\mathbf{r}_1$ through a telescope while the idler is directly sent to a local photodetector D$_2$ at $\mathbf{r}_2 \simeq 0$, kept isolated from environmental effects such as turbulence. The reflected signal photon is received by the telescope and is annihilated at photodetector D$_1$, which is placed at the focal point of the telescope. The registration times of the signal photon and the idler photon, $t_1$ and $t_2$, respectively, are recorded by two event timers whose time bases are synchronized by a clock.  The joint photodetection events at different values of $t_1 - t_2$ are counted and analyzed by a coincidence circuit. The circuit produces a histogram of number of coincidences vs. different values of $t_1 - t_2$. Due to the broad spectrum of the SPDC, the histogram is usually a sharp, narrow function of $t_1 - t_2$ with its maximum value at $t_1 - t_2 = (r_1 - r_2)/c$, where $r_1$ is the total optical distance from the biphoton source to the target and finally to D$_1$, $r_2$ is the optical distance between the biphoton source and D$_2$; while the single-photon counting rates of D$_1$ and D$_2$ are always constant. The position of the target is then easily evaluated from the histogram. Interestingly, this measurement is turbulence-resistant.

Before presenting the details of how this scheme is realized experimentally, we first discuss the theory behind this proof-of-principle Lidar system from the perspective of \textit{nonlocal two-photon interference}, mainly focusing on two  questions: (1) Why is the measured histogram a narrow, sharp function of $t_1 - t_2$ while the pump, signal and idler are all CW fields? (2) Why is the histogram turbulence-resistant? In 1935, Einstein, Podolsky and Rosen (EPR) proposed an entangled two-particle system \cite{EPR}. In such a system, wherein the total momenta of the two particles are conserved, the particles propagate in opposite directions to distant locations. Although particle-1 and particle-2 can be individually observed at any distant positions; due to the nonlocal two-photon interference, i.e, coherent superposition of a large number of two-photon amplitudes of the entangled particle pair,  if particle-1 is observed at a certain position $x_1$, particle-2 must be observed at a unique position $x_2$ jointly and vice-versa, corresponding to a narrow, sharp correlation function of $x_1 - x_2$.  The reported quantum Lidar system is a realization of the temporal manifestation of this EPR correlation. According to quantum theory of light, the probability of having a joint photodetection event at space-time points $(\mathbf{r}_1, t_1)$ and $(\mathbf{r}_2, t_2)$ is proportional to the second-order coherence function or correlation function of the measured fields \citep{Glauber, ScullyBook}:

\begin{align}\label{eqn:CorrFunc1}
& G^{(2)}(\mathbf{r}_1, t_1;\mathbf{r}_2, t_2) \cr
& = \Big\langle \Psi \big| \hat{E}^{(-)}(\mathbf{r}_1, t_1)\hat{E}^{(-)}(\mathbf{r}_2, t_2) 
\hat{E}^{(+)}(\mathbf{r}_2, t_2) \hat{E}^{(+)}(\mathbf{r}_1, t_1) \big| \Psi \Big\rangle \cr
\end{align}
where $\hat{E}^{(-)}(\mathbf{r}_j, t_j)$ and $\hat{E}^{(+)}(\mathbf{r}_j, t_j)$, $j = 1,2$, are the negative-frequency and the positive-frequency field operators, respectively, at space-time point $(\mathbf{r}_j, t_j)$. 

For the two-photon entangled state of SPDC, $G^{(2)}(\mathbf{r}_1,t_1;\mathbf{r}_2,t_2)$ can be written as the modulus square of a two-photon effective wavefunction, or biphoton \cite{Rubin}
\begin{align}\label{eqn:CorrFunc2}
& G^{(2)}(\mathbf{r}_1,t_1;\mathbf{r}_2,t_2) \cr
& = \Big{|} \big\langle 0 |\hat{E}^{(+)}(\mathbf{r}_2, t_2) \hat{E}^{(+)}(\mathbf{r}_1, t_1) |\Psi \big\rangle \Big{|}^2  
\equiv \big{|} \Psi(\mathbf{r}_1,t_1;\mathbf{r}_2,t_2) \big{|}^2 \cr
\end{align}
where the kets $| 0 \rangle$ and $|\Psi \rangle$ stand respectively for the vacuum and the state of the signal-idler photon pair.

Consider a simplified 1-D experimental setup by assuming that the signal and the idler are emitted from a point-like source and are jointly detected by two point-like photon counting detectors D$_1$ and D$_2$, respectively. To calculate the temporal (longitudinal) two-photon correlation function $G^{(2)}(\tau_1, \tau_2)$, where $\tau_j \equiv t_j - r_j/c$, $j = 1,2$, the two-photon state of the signal-idler photon pair can be approximated as
\begin{align} \label{eqn:SignalIdler}
\left| \Psi \right\rangle \simeq \Psi_0 \, \sum_{s,i}\delta \left( \omega
_{s}+\omega _{i}-\omega _{p}\right) a_{s}^{\dagger }(\omega_{s})\, a_{i}^{\dagger }(\omega_{i})\mid 0\rangle 
\end{align} 
Here, the subscripts \textit{s}, \textit{i} and \textit{p} denote the signal, the idler, and the pump, respectively. The two-photon effective wavefunction of the biphoton $\Psi(\tau_1, \tau_2)$ is calculated as \cite{Valencia}
\begin{align}\label{eqn:Biphoton1}
\Psi(\tau_1, \tau_2) 
& \simeq \Psi_0 \sum_{s,i}\delta \left( \omega_{s}+\omega _{i}-\omega _{p}\right) e^{-i\omega_s \tau_{1}} e^{-i\omega_i \tau_{2}} \cr
& = \Psi_0 \sum_{s,i} \Psi_{s,i}(\omega_s, \omega_i)
\end{align}
representing a coherent superposition of a large number of two-photon amplitudes. This superposition indicates a nonlocal two-photon interference: a pair of entangled photons separated by a distance interfering with the pair itself \cite{DiracQM, Note}. It is this two-photon interference resulting in a narrow, sharp histogram from the measurement of CW waves:
\begin{align}\label{eqn:Biphoton2}
& \Psi (\tau_{1}; \tau_{2}) \cr
&\cong \Psi_0 \, e^{-i (\omega^{0}_{s} \tau_{1} + \omega^{0}_{i} \tau_{2})} \int d \nu \, f(\nu) \, e^{-i \nu (\tau_{1} -\tau_{2})} \cr
&= \Big\{\Psi_{0} \, e^{-\frac{i}{2} (\omega^{0}_{s} + \omega^{0}_{i}) (\tau_{1} + \tau_{2})} \Big\} \cr 
& \ \ \ \ \ \times \Big\{ {\mathcal F}_{\tau_{1}-\tau_{2}} \big[ f(\nu)\big] e^{-\frac{i}{2} (\omega_{s}^{0}- \omega_{i}^{0})(\tau_{1} - \tau_{2})}\Big\}
\end{align}
where $\omega^0_s$ and $\omega^0_i$ are the central frequencies of the signal and idler, respectively; $\nu \equiv \omega_s - \omega^0_s$ is the detuning of the signal field, ${\mathcal F}_{\tau_{1}-\tau_{2}} \big\{ f(\nu)\big\}$ is the Fourier transform of the spectral function of the signal-idler field. The temporal (longitudinal) correlation function $G^{(2)}(\tau_1- \tau_2)$ is thus:
\begin{align}\label{eqn:CorrFunc3}
G^{(2)}(\tau_1- \tau_2) \propto \Big| \, {\mathcal F}_{\tau_{1}-\tau_{2}} \big[ f(\nu)\big] \Big|^{2}
\end{align}
The Fourier transform is usually a $\delta$-function-like function of $\tau_1 - \tau_2$ in the case of SPDC due to its wide spectrum.  Assuming a constant distribution of $\nu$ within a certain spectrum $\Delta \omega$, the Fourier transform can be approximated as a sinc-function
\begin{align}\label{eqn:CorrFunc4}
G^{(2)}(\tau_1- \tau_2) \propto \textrm{sinc}^2 \Big[\frac{\Delta \omega (\tau_1 - \tau_2)}{2\pi}\Big]. 
\end{align}
The sinc-function becomes a $\delta$-function-like narrow and sharp function of $\tau_1 - \tau_2$ when $\Delta \omega \sim \infty$.  At $\tau_1 - \tau_2 = 0$, i.e., $t_1 - t_2 = (r_1 - r_2)/c$ the sinc-function takes its maximum value of one.

Indeed, it is the nonlocal two-photon interference resulting in the narrow, sharp histogram. In the language of Einstein-Podolsky-Rosen: the signal photon and the idler photon are observable by D$_1$ and D$_2$, respectively, at any time; however, if the signal photon is observed by D$_1$ at $t_1$, the idler photon must be jointly observed by D$_2$ at a unique time $t_2$ \cite{EPR}.

From this two-photon superposition or two-photon interference, it is not difficult to find that if all two-photon amplitudes $\Psi_{s,i}(\omega_s, \omega_i)$ of the signal-idler experience the same turbulence along the optical path with the same phase variations, 
\begin{align}\label{eqn:Turb1}
& \Psi^{T}(\tau_1, \tau_2) \cr
& \simeq \Psi_0 \sum_{s,i}\delta \left( \omega_{s}+\omega _{i}-\omega _{p}\right) \Big\{e^{-i\delta \varphi(\mathbf{r}_1)}\Big\} e^{-i\omega_s \tau_{1}} e^{-i\omega_i \tau_{2}} \cr
& = \Big\{ \Psi_0 \, e^{-i\delta \varphi(\mathbf{r}_1)} \Big\} \Psi(\tau_1, \tau_2)
\end{align}
the nonlocal two-photon correlation function 
\begin{align} \label{eqn:Turb2}
G^{(2)}(\tau_1- \tau_2) = \big|\Psi^T(\tau_1, \tau_2)\big|^2 = \big|\Psi(\tau_1, \tau_2)\big|^2
\end{align}
will be invariant. The reported quantum Lidar system has taken advantage of this two-photon interference: a signal-idler photon pair has many different yet indistinguishable amplitudes to produce a joint photodetection event. The superposed two-photon amplitudes experience the same turbulence along the optical path with the same phase variations. The measured joint detection histogram is therefore turbulence-resistant.

In addition to the turbulence-resistant property analyzed above, the correlation measurement is practically unaffected by background noise of thermal light: the thermal light noise has very small chances to contribute to the coincidence events of D$_1$ and D$_2$, especially when the noise is applied to D$_1$ only while D$_2$ is kept isolated to minimize the effects of the surrounding environment. This additional noise-resistant demonstration is similar to  that reported by Rarity \emph{et. al}. in their 1990 quantum ranging experiment \citep{RarityNoise}.

\begin{figure}[H]
\centering
\includegraphics[width=1.0\linewidth]{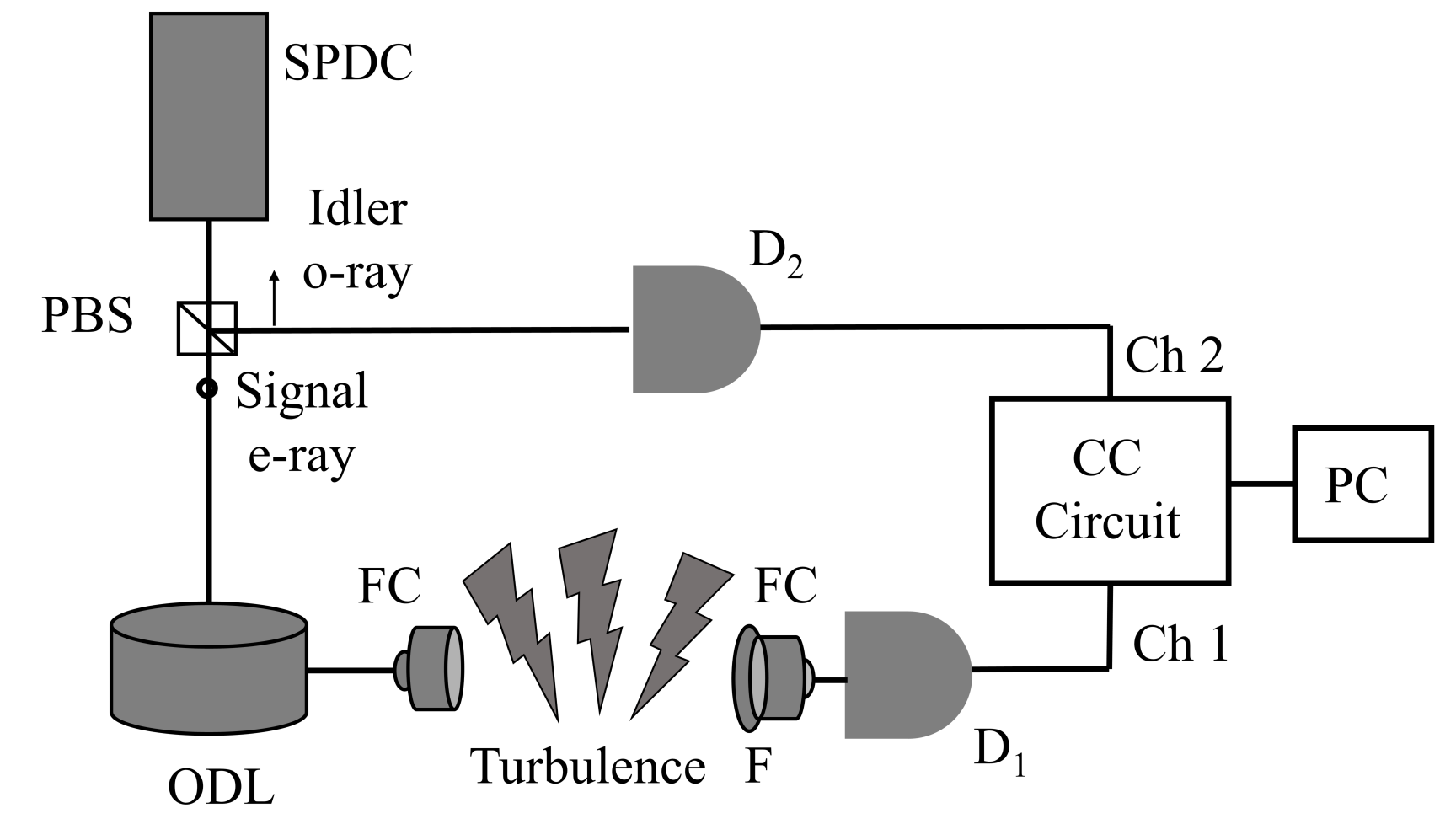}
\caption{Proof-of-principle correlation measurement setup for EPR pairs subjected to turbulence. Relevant abbreviations: Spontaneous Parametric Down-conversion (SPDC), Polarizing Beam Splitter (PBS), Optical Delay Line (ODL), Fiber Collimator (FC), Spectral Filter (F), Coincidence Counting (CC) Circuit.} 
\label{fig:setup}
\end{figure}

\begin{figure*}[ht]
\begin{minipage}[t]{1.0\textwidth}
\begin{tabular}{c|c}
\subfloat[]{\includegraphics[width = 0.49\linewidth]{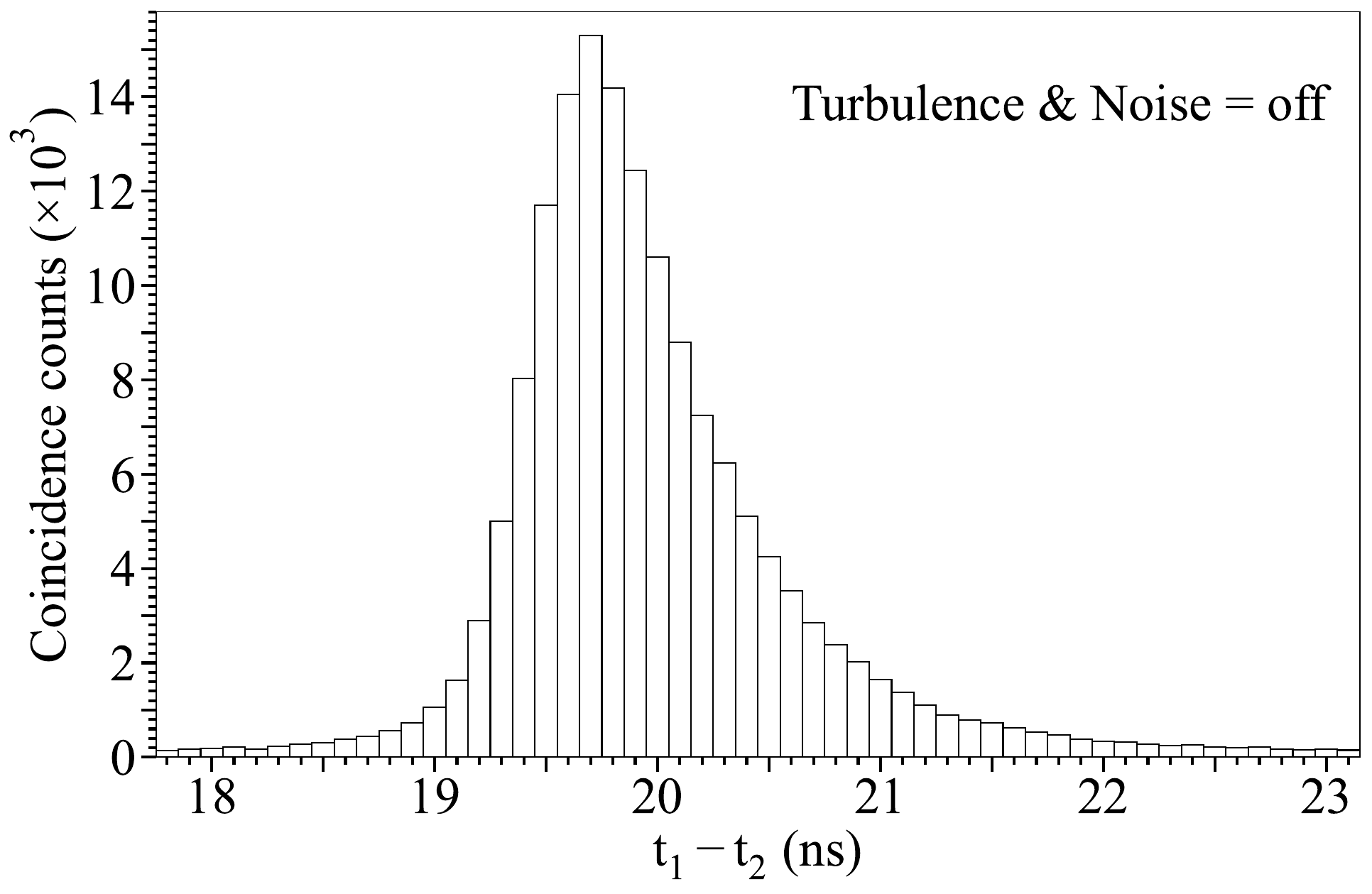}} &
\subfloat[]{\includegraphics[width = 0.49\linewidth]{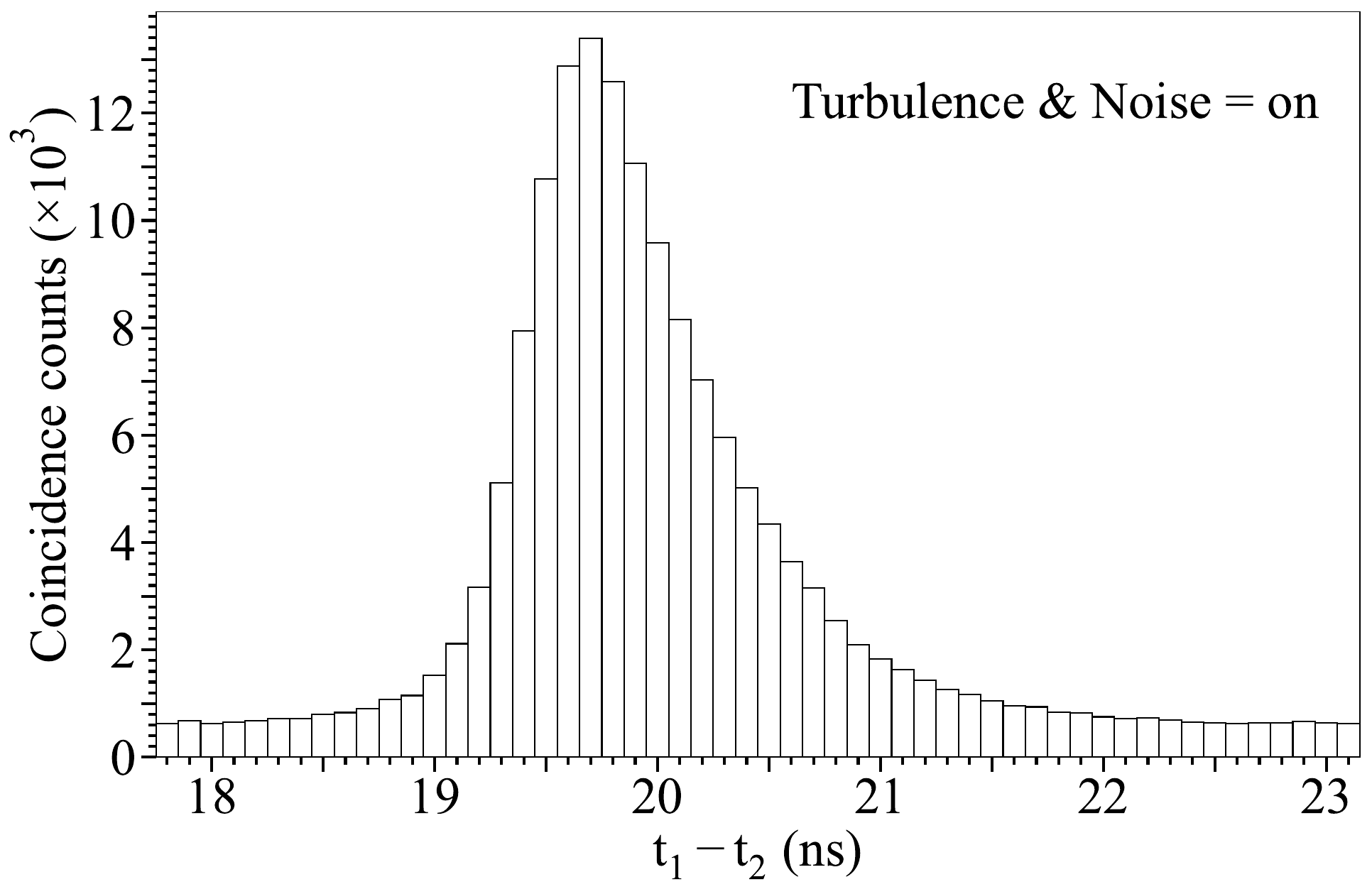}} 
\end{tabular}
\hrule
\bigskip{}
\centering
\subfloat[]{\includegraphics[width = 0.5\linewidth]{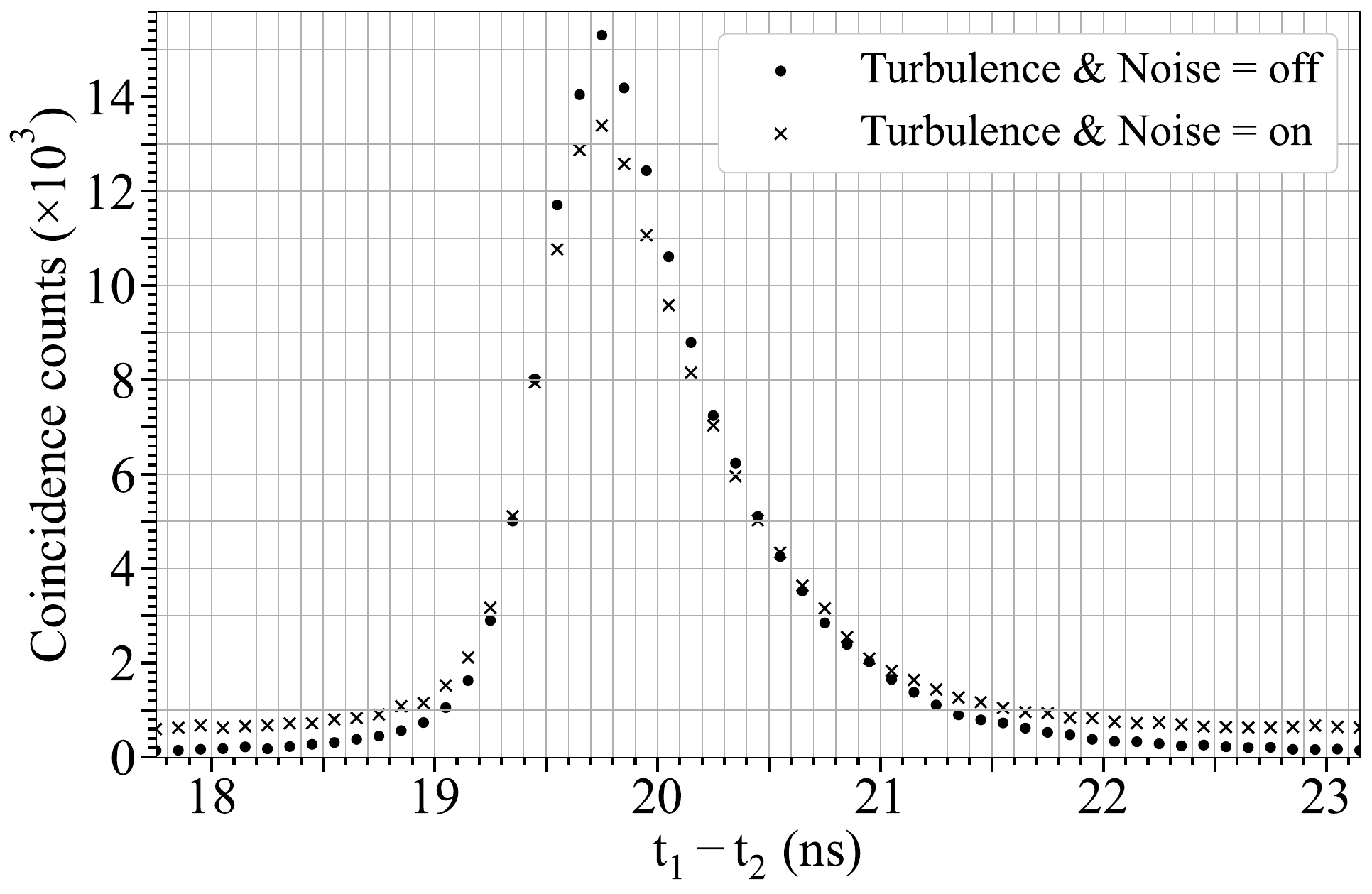}}
\hrule
\end{minipage}
\caption{Typical measured histograms: number of coincidence counts vs. temporal delay ($t_1-t_2$) in nanoseconds. (a) Without turbulence and noise; (b) With significant level of turbulence and noise; (c) Data for both histograms displayed in a single plot for comparison.}
\label{fig:turbulence}
\end{figure*}

The proof-of-principle experimental setup is shown in Fig.~\ref{fig:setup}. An entangled biphoton source of degenerate Type-II SPDC is used to generate orthogonally polarized signal-idler photon pairs. Roughly speaking, the process of SPDC involves sending a pump laser beam into a nonlinear material, such as a non-centrosymmetric crystal. Occasionally, the nonlinear interaction inside the crystal leads to the annihilation of a high frequency pump photon and the creation of a pair of entangled lower frequency photons as signal and idler, namely a ``biphoton''. In our experiment, the pump is a 405 nm single-mode CW laser. The central wavelengths of the signal photon and the idler photon are at 810 nm. A polarizing beam-splitter (PBS) is used to separate the signal and the idler. Optical fibers with adjustable lengths are used to simulate the time delay of distant targets. The idler is coupled directly into a local point-like single-photon counting avalanche photodiode (SPAD) D$_2$, while the signal is first coupled into two pieces of optical fibers optimized for wavelengths above 800 nm, and then coupled to another point-like SPAD D$_1$. The two fibers are optically coupled but separated in free space (air) by a distance of $\sim1$m. A high power toaster oven blows hot air through the  free path between the two fibers to produce turbulence in the entire $\sim1$m free space (air) path. The introduced turbulence, which aims to mimic atmospheric turbulence, is strong enough to blur out the interference pattern of a double-slit interferometer. The coincidence counting circuit records the registration times of the photodetection events of D$_1$ and D$_2$, counts and analyzes their joint-detection events at different values of $t_1 - t_2$. A coincidence histogram is then generated from the output data of the circuit. 

Different lengths of optical fiber delays, ranging from 0 to 1000 m, are used to simulate the time delays throughout our measurements for testing various experimental parameters. The integration time for the measurements is chosen in such a way that it is long enough as compared to the timescale of the variation of strong turbulence. In our additional tests for the system performance under noise, the background noise is made about 20 times greater than the signal level (limited by the damage threshold of the photon counting detectors). Two typical histograms for turbulence-resistant and noise-free measurements and an additional plot showing both measurements for comparison are reported in Fig.~\ref{fig:turbulence}. Comparing the two histograms, one without turbulence and noise, and the other with significant levels of both turbulence and noise, no discernible change in either the position of the peak or the function-width of the histogram is found, except for a slightly enhanced baseline caused by the background noise, and a slightly decreased peak count due to the scattering loss caused by the applied atmospheric turbulence. It should also be noted that both sets of measurement presented here achieved sub-centimeter positioning resolution, determined by the resolution of the coincidence circuit.

In summary, we have demonstrated the working mechanism of a turbulence-resistant quantum Lidar system. The precision of optical timing and positioning measurements based on classic Lidar systems utilizing classical light pulses can suffer from the undesirable effects of atmospheric turbulence. As demonstrated by this work, the two-photon interference of an EPR state, in which the timing for the detection of a photon ensures a unique timing for the detection of the other photon when they are jointly detected, is seen to be robust against such effects and can be exploited as the basis for quantum Lidar applications. The observed invariant two-photon correlation function suggests a potential quantum technology for diverse applications such as high precision positioning-timing, rangefinding, quantum sensing, metrology and imaging Lidar system under turbulence and noise.

\begin{acknowledgments}
We would like to thank T. A. Smith and M. F. Locke for helpful discussions. This work was partially supported by Northrop Grumman Corporation. 
\end{acknowledgments}

\section*{Author Declarations}
\subsection*{Conflict of Interest}
The authors have no conflicts to disclose.
\subsection*{Author Contributions}
\textbf{Binod Joshi}: Investigation (lead); Data Curation (lead); Formal Analysis (equal); Writing - original draft (equal). \textbf{Michael M. Fitelson}: Funding acquisition (lead); Conceptualization (equal). \textbf{Yanhua Shih}: Supervision (lead); Conceptualization (equal); Formal Analysis (equal); Writing - original draft (equal).
\section*{Data Availability}
The data that support the findings of this study are openly available in Zenodo at \url{https://zenodo.org/doi/10.5281/zenodo.10536057}.

\bibliography{ref_combined}

\begin{thebibliography}{36}
\bibitem [{Not()}]{Note}%
  \BibitemOpen
  \href@noop {} {}\bibinfo {note} {{T}o avoid any misunderstanding, we would like to emphasize that (1) two photon interference is not the interference between two photons. In Dirac’s language: a pair of photons only interferes with the pair itself. (2) Two-photon interference is not limited by ``two-term'' superposition. The superposition of a large number of two-photon amplitudes is a typical quantum interference phenomenon.}\BibitemShut {Stop}%
\end{thebibliography}

\end{document}